%% file: main_arXiv.tex
\documentclass[10pt, conference, letterpaper]{ IEEEtran}
\IEEEoverridecommandlockouts

\usepackage{cite}
\usepackage{amsmath,amssymb,amsfonts}
\usepackage{algorithmic}
\usepackage{graphicx}
\usepackage{textcomp}
\usepackage{xcolor}

\usepackage{algorithmic}
\usepackage{algorithm}

\newcommand{\squishlist}{
\begin{list}{$\bullet$}{
  \setlength{\itemsep}{0pt}
  \setlength{\parsep}{3pt}
  \setlength{\topsep}{3pt}
  \setlength{\partopsep}{0pt}
  \setlength{\leftmargin}{3.5mm}
  \setlength{\labelwidth}{1em}
  \setlength{\labelsep}{0.5em}}}
\newcommand{\squishend}{\end{list}}

\usepackage{subfig}
\usepackage{xspace}
\usepackage{textcomp}
\usepackage{booktabs}
\usepackage{url}
\usepackage{hyperref}
\usepackage{xcolor}
\usepackage{mfirstuc}
\usepackage{balance}
\usepackage{marvosym}
\usepackage{diagbox}
\usepackage{comment}
\usepackage{threeparttable}

\def\BibTeX{{\rm B\kern-.05em{\sc i\kern-.025em b}\kern-.08em
    T\kern-.1667em\lower.7ex\hbox{E}\kern-.125emX}}
\begin{document}

\title{
iPanda: An LLM-based Agent for Automated Conformance Testing of Communication Protocols
}
\author{
\IEEEauthorblockN{
Xikai Sun\IEEEauthorrefmark{1}, 
Fan Dang\IEEEauthorrefmark{2}\textsuperscript{\Letter}, 
Shiqi Jiang\IEEEauthorrefmark{3},
Jingao Xu\IEEEauthorrefmark{4},
Kebin Liu\IEEEauthorrefmark{1},
Xin Miao\IEEEauthorrefmark{1},
Zihao Yang\IEEEauthorrefmark{5}, }
\IEEEauthorblockN{
Weichen Zhang\IEEEauthorrefmark{1}, 
Haimo Lu\IEEEauthorrefmark{1}, 
Yawen Zheng\IEEEauthorrefmark{1},
Yunhao Liu\IEEEauthorrefmark{1}\textsuperscript{\Letter}
}

\IEEEauthorblockA{
\IEEEauthorrefmark{1}Tsinghua University     
\IEEEauthorrefmark{2}Beijing Jiaotong Universi
\IEEEauthorrefmark{3}Microsoft Research Asia}
\IEEEauthorblockA{
\IEEEauthorrefmark{4}Carnegie Mellon University
\IEEEauthorrefmark{5}Yanshan University
$^\textrm{\Letter}$ Corresponding Authors
}
}
\maketitle

\begin{abstract}
Conformance testing is essential for ensuring that protocol implementations comply with their specifications. However, traditional testing approaches involve manually creating numerous test cases and scripts, making the process labor-intensive and inefficient. Recently, Large Language Models (LLMs) have demonstrated impressive text comprehension and code generation abilities, providing promising opportunities for automation. In this paper, we propose iPanda, the first framework that leverages LLMs to automate protocol conformance testing. Given a protocol specification document and its implementation, iPanda first employs a keyword-based method to automatically generate comprehensive test cases.
Then, it utilizes retrieval-augmented generation and customized CoT strategy to effectively interpret the implementation and produce executable test programs.
To further enhance programs' quality, iPanda incorporates an iterative optimization mechanism to refine generated test scripts interactively. Finally, by executing and analyzing the generated tests, iPanda systematically verifies compliance between implementations and protocol specifications. Comprehensive experiments on various protocols show that iPanda significantly outperforms pure LLM-based approaches, improving the success rate ($Pass@1$) of test-program generation by factors ranging from $4.675\times$ to $10.751\times$.
\end{abstract}


\input{1_introduction}
\input{2_background}

\input{3_design}
\input{4_experiment}
\input{5_relatedWork}
\input{6_conclusion}

\bibliographystyle{IEEEtran}
\bibliography{IEEEabrv, ref}

\end{document}

%% file: 1_introduction.tex
\section{Introduction}
\label{sec:1}

Communication protocols such as MQTT, CoAP, WebSocket form the backbone of today's information-driven society, playing critical roles in internet data transmission, IoT connectivity, and cloud computing services. Ensuring the stability and efficiency of protocol implementations is therefore paramount and necessitates rigorous testing procedures.
Conformance testing, specifically designed to verify adherence of protocol implementations to their official standards and specifications, is one of the essential methods employed in validating communication protocols. 
In typical scenarios, testers might spend weeks manually writing extensive test scripts utilizing protocol implementation libraries, and individually verifying compliance with protocol requirements. 
For instance, a typical compliance check for the CoAP protocol involves developing hundreds of script-based tests, to verify its message grammar, stateful interactions and operational sequences, as mandated by over $200$ \texttt{MUST} and \texttt{SHOULD} requirements in RFC7252. 
With the increasing complexity of protocols, traditional manual testing methods have become inefficient, cumbersome, and difficult to generalize. This situation underscores an urgent need for automated testing solutions that minimize human intervention while enhancing efficiency and coverage.
 
Concurrently, large language models (LLMs) have shown impressive capabilities in language comprehension, reasoning, and performing sophisticated tasks. Trained on extensive datasets, LLMs effectively generalize to novel tasks, parse complex inputs, and maintain context, making them ideal for automating intricate operations. For instance, in IoT fuzz testing, LLM-driven automation enhances protocol message generation, increasing vulnerability detection effectiveness and uncovering previously undetected issues~\cite{p_llmif_wang2024llmif}. Similarly, in mobile device automation, LLMs combine general reasoning with domain-specific expertise, facilitating complex tasks without extensive manual scripting~\cite{p_autodroid_wen2024autodroid, p_autodroidv2_wen2024autodroid}. These examples demonstrate the transformative potential of LLMs in turning labor-intensive procedures into efficient automated solutions.
  
Given this context, a compelling question arises: Can we leverage the capabilities of LLMs to address the challenges inherent in conformance testing of communication protocols? Realizing this vision in practice, however, entails overcoming several key obstacles:
\squishlist
    \item \textbf{Specification-to-test-case derivation}. In contrast to fuzzing, conformance testing is intrinsically driven by specifications, requiring the translation of often ambiguous textual specifications into structured, executable test logic.
    \item \textbf{Knowledge adaptation and grounding}. LLMs struggle to bridge the gap between abstract rules described in test cases and the concrete APIs provided by specific implementation libraries, making it difficult to generate valid and context-aware test programs. 
    \item \textbf{Practical executability and reliability}. Beyond syntactic correctness, test programs must be semantically aligned with the target library's API to ensure they are both executable and trustworthy in real-world testing scenarios. 
\squishend
Addressing these challenges is essential not only to resolve the immediate practical issues but also to establish foundational methodologies as LLM technologies continue to advance in network protocol testing.

While LLMs have achieved remarkable success in domains such as code generation and dialogue systems, they fall short in addressing the unique demands of protocol conformance testing.
We bridge this gap with \textbf{iPanda}, an \textbf{I}ntelligent \textbf{P}rotocol Testing \textbf{an}d \textbf{D}ebugging \textbf{A}gent.  
As the first LLM-based solution for automating communication protocol conformance testing, iPanda streamlines the entire workflow. It autonomously parses a protocol specification to generate comprehensive test cases, translates them into executable programs for a target implementation library, and then executes these tests to systematically identify compliance violations, thereby significantly reducing manual effort.


Especially, we introduce multiple mechanisms in iPanda to overcome the above challenges. 
To efficiently generate test cases aligned with precise testing requirements, we propose a novel keyword-based test-case generation approach grounded in the inherent characteristics of protocol specifications. It automatically locates the specific specifications that need to be tested based on keywords, and generates formatted test cases.
\color{black}
Moreover, to ensure the adaptation of iPanda to varying protocol libraries, we introduce code-oriented retrieval-augmented generation (RAG), which is deeply coupled with the test program generation process. This mechanism iteratively queries the knowledge base at different stages of program synthesis, providing just-in-time, context-specific guidance. 
Additionally, to enable LLMs to handle complex test cases, we design a customized chain-of-thought (CoT) strategy that operationalizes a divide-and-conquer methodology.
It guides iPanda to first decompose the complex task case into logical sub-tasks and solve them individually, before integrating these modular solutions into a complete, high-quality test program.
Finally, We introduce testing platform and iterative optimization mechanism to achieve automated testing and program correction, ensuring the program's executability and reliability.
\color{black}

iPanda also supports natural-language commands and demonstrates reasoning capabilities for complex state transitions, making it effective for dynamic, heterogeneous network environments where traditional methods frequently fall short. By significantly reducing the dependency on human expertise and fully automating test execution, iPanda presents a transformative advancement in protocol conformance testing.

Our contributions are as follows:
\squishlist
    \item To the best of our knowledge, 
    we present the first automated protocol conformance testing framework integrating LLMs with domain-specific expertise, 
    including keyword-based test case generation, customized CoT strategy, and iterative optimization mechanism for test-program refinement.
    \item We design and implement \textbf{iPanda}, an intelligent, LLM-powered agent capable of autonomously extracting test cases from protocol specifications, invoking protocol implementation libraries, executing and debugging tests, and identifying conformance issues.
    \item \color{black} As an example, we using iPanda to perform conformance testing on CoAP and RSocket, and present two test case sets, i.e., CoAP-set and RSocket-set. \color{black} 
    Comprehensive experiments demonstrate that, compared to the pure LLM-based method, iPanda improves the $Pass@1$ of test program generation by $4.675 \times$ to $10.751 \times$.
\squishend

Upon acceptance of this paper, we will publicly release the associated source code and datasets.

%% file: 2_background.tex
\section{Background and motivation}
\label{sec:2}

\subsection{Conformance Testing of Protocols}
\label{sec:2.1}
Conformance testing for communication protocols is a method used to verify whether a protocol implementation complies with the requirements defined by the protocol specification. 
Its primary goal is to ensure that the protocol implementation behaves in a conformant manner across various scenarios and input conditions, thereby guaranteeing compatibility and interoperability between different protocol implementations.
A complete protocol conformance testing process typically includes the following key steps:  
\squishlist
    \item \textbf{Specification analysis}. 
    Thoroughly understand and analyze the protocol's standard documents to extract explicitly defined behaviors and requirements.
    \item \textbf{Test case design}. 
    Develop comprehensive test cases based on the protocol specifications.
    \item \textbf{Test case implementation}.
    Write test scripts or programs using the protocol implementation under test to concretely realize the test cases. 
    \item \textbf{Test execution}. 
    Run the test case programs in a properly configured testing environment and collect results.
    \item \textbf{Result analysis and evaluation}.
    Analyze the test results to determine whether the system under test complies with the specified requirements.
\squishend

\color{black}
\subsection{Uniqueness of Conformance Testing}
Traditional testing, like unit testing and integration testing, is fundamentally implementation-driven.
Whether verifying individual code components (unit testing) or the interactions between modules (integration testing), the design, execution, and evaluation of test cases are all rooted in the implementation’s source code and architecture. 
In short, it aims to ensure that the existing implementation works as expected.
In contrast, specification-based conformance testing differs fundamentally in its methodology. 
Its goal is to verify whether an implementation strictly adheres to an external, authoritative specification (e.g., a technical standard or an official API document). 
Its uniqueness is reflected mainly in two aspects:
\squishlist
    \item \textbf{Shift in the source of truth.}
    In conformance testing, the source of truth shifts from the internal code to the external specification. 
    It does not concern itself with implementation details but focuses solely on whether the behavior aligns with what is described in the specification.
    \item \textbf{Systematic specification-driven process.} 
    Conformance testing requires a top-down process. 
    It begins with systematically interpreting and decomposing the specification text into verifiable requirements. 
    Based on these extracted requirements, test cases are then generated and executed to map precisely to actual API calls.
\squishend

Essentially, traditional testing verifies what the code does, whereas conformance testing verifies whether the code does what the specification requires it to do. This makes conformance testing a systematic endeavor that demands holistic understanding (inferring the functional points to be tested from the specification), careful planning (designing tests that can validate these points), and rigorous execution (generating test programs to implement these tests). It is not merely a straightforward check of code functionality. For this reason, traditional testing tools are inadequate for conformance testing.
\color{black}

\subsection{Integrating LLM and Conformance Testing}
\label{sec:2.3}

The current conformance testing process has significant limitations, primarily due to its heavy reliance on manual effort. 
It involves a vast number of detailed test cases that must be manually written, making the process labor-intensive and time-consuming. 
Additionally, implementing these test cases requires extensive test program development, further increasing the cost of development and maintenance.
These limitations significantly reduce the efficiency and flexibility of conformance testing.  

At present, no conformance testing tool exists that seamlessly integrates protocol documents, implementations, and test results. 
The exceptional text comprehension and code generation capabilities of LLMs offer new possibilities for addressing these shortcomings. 
By leveraging the powerful abilities of LLMs, test case design and test program development can be automated or semi-automated, substantially reducing the manual workload.
Therefore, this work aims to integrate LLMs into the conformance testing process for communication protocols. 
We propose an LLM-based conformance testing agent framework to enhance the efficiency and effectiveness of conformance testing.

%% file: 3_design.tex
\section{Design of iPanda}
\label{sec:3}
In this section, we present the design of iPanda in detail, which is specifically designed for conformance testing of communication protocol.
iPanda can dynamically generate test case sets for conformance testing based on protocol specification documents.
For a given protocol implementation library, it can generate test case programs, interact with the implementation in a simulated communication environment, perform testing and debugging, and analyze execution results to identify potential deficiencies in the implementation.


\begin{figure}[tbp]
    \centering{\includegraphics[width=1\columnwidth]{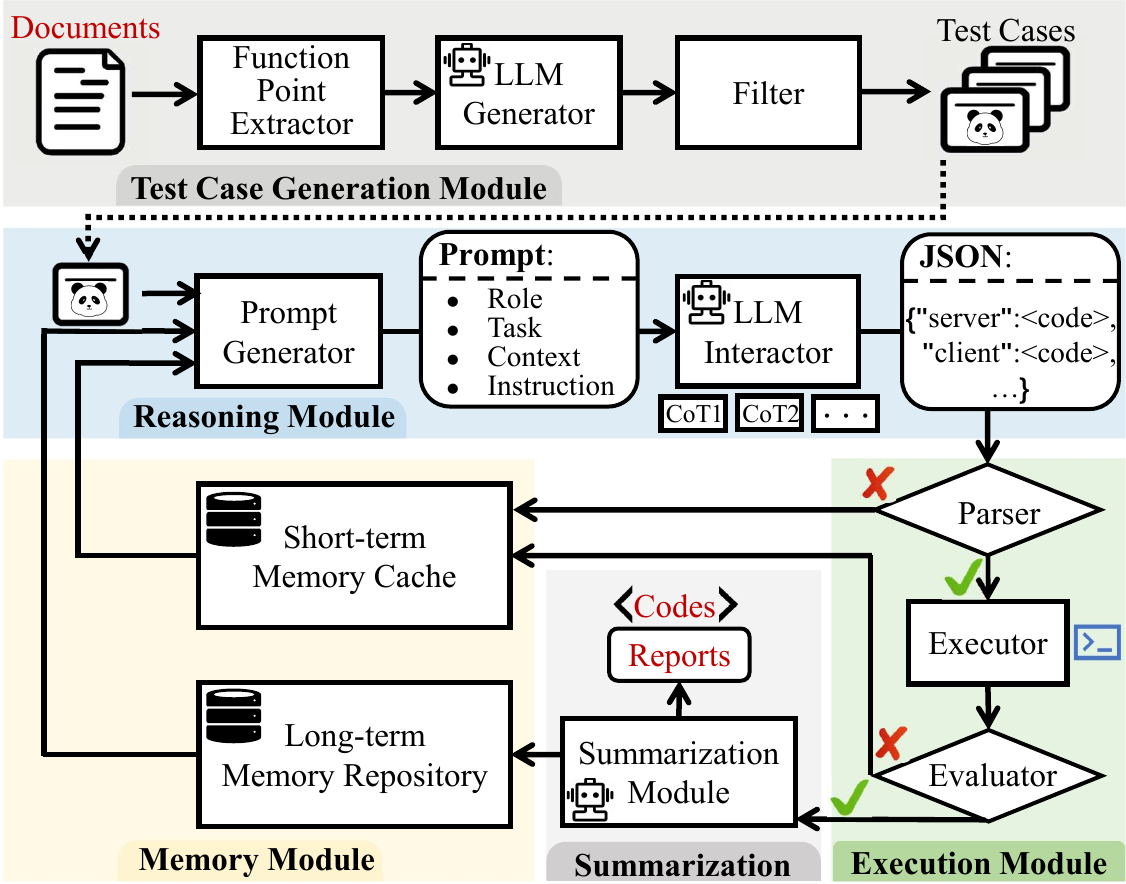}}
    \caption{The overview of iPanda.}
    \label{fig:overview}
\end{figure}

The overview of iPanda is shown in Fig.~\ref{fig:overview}. 
Taking the RSocket protocol~\cite{p_rsocket} as an example, suppose a user wants to verify whether its Python implementation \texttt{rsocket-py}~\cite{p_rsocket-py} conforms to its specification.  
iPanda first extracts key functional points from the protocol document and automatically generates standardized test cases using the LLM generator, optionally applying a filter to remove anomalous cases. For each test case, guiding by customized CoT strategy, iPanda generates executable test program using the target implementation library. To generate high-quality program, it retrieves relevant context from the implementation library. To ensure executability, the generated program undergoes validation; if issues are detected, iPanda initiates iterative refinement using historical context and error information, dynamically adjusting prompts to debug the program. Once the generation yields executable program or reaches a retry limit, iPanda compiles a final debugging report and evaluates test outcomes, determining compliance with the protocol’s conformance requirements.

To address the challenges mentioned in Sec.~\ref{sec:1} and enhance iPanda's effectiveness, we have designed several optimization methods.
The following sub-sections will provide detailed descriptions of these methods.

\subsection{Keyword-based Test Case Generation}
\label{sec:3.1}
\color{black}
In conformance testing, the generation of test cases has traditionally relied heavily on manual effort. 
For example, in the FIDO2 Conformance Test Tool for the FIDO protocol, developers manually extract as many test requirements as possible from the specification documents and then write corresponding test cases to integrate into the tool\cite{p_fido2_rammatopoulos2022blind}. 
Although some efforts have emerged to automate the parsing of specifications (For instance, RFCSCAN attempts to treat individual RFC sections as test units and directly compare them with the implementation\cite{p_rfcscan_zheng2025llm}), these approaches remain limited to coarse-grained, section-level parsing. 
As a result, they lack the capability for more fine-grained analysis of the documents, making it challenging to automatically generate test cases targeting specific functional points.
\color{black}

\textbf{Insight}: The majority of protocol documents follow strict format requirements and are structured using specific terminology to define and organize key concepts.

Taking CoAP as an example, we reviewed a total of $32$ RFC documents related to CoAP and found that $87.5\%$ of them comply with RFC 2119~\cite{p_coap_rfc7252_shelby2014rfc, p_coap_32}. 
These documents use uppercase keywords such as \texttt{MUST}, \texttt{REQUIRED}, \texttt{SHALL}, to define and emphasize critical protocol specifications and requirements (named the functional points). 
Even the documents that do not strictly follow RFC 2119 still use similar auxiliary verbs as keywords  to highlight protocol functional points. 

\color{black}
\textbf{Preliminary Experiment}: Furthermore, we examine the coverage when using keywords to locate functional points. Assuming the keyword set is $K$, and following RFCSCAN, we take the smallest subsections $s_i, i=1,...,N$ as the evaluation units. The section coverage $C$ is then defined as follows:
\begin{equation}
    C=\frac{\sum_{i=1}^NL(s_i)\cdot I(s_i)}{\sum_{i=1}^NL(s_i)},
\end{equation}
where the function $L(s_i)$ measures the string length of $s_i$, since sections vary in length and longer sections are more likely to contain additional functional points. $I(s_i)$ is the indicator function, defined as follows:
\begin{equation}
    I(s_i) = 
    \begin{cases}
        1, & \exists k \in K, k \in s_i, \\
        0, & \text {otherwise}. 
    \end{cases}
\end{equation}
The experiment shows that for 32 CoAP-related RFC documents (considering only the main body), the overall section coverage reaches $87.50\%$, with 10 documents achieving a section coverage of $100\%$. These results demonstrate that using keywords to locate functional points can parse documents at a finer granularity while ensuring high section coverage.
\color{black}

Inspired by the above insight and experiment, we design a novel specification-driven test case generation method named keyword-based test case generation (keyword-based TCG). 
This method is integrated into the test case generation module, as shown in Fig.~\ref{fig:overview}. 
Keyword-based TCG combines heuristic rules with the idea of generating datasets using LLMs. 
Specifically, this method first detects whether the document follows RFC 2119. 
If it does, keyword-based TCG uses regular expressions to extract paragraphs containing uppercase keywords; if not, it
defaults to case-insensitive keyword extraction.
Each extracted paragraph is a complete natural paragraph to preserve as much contextual semantic information as possible. 
We define these paragraphs as functional points. 

\begin{figure}[tbp]
    \centering{\includegraphics[width=0.9\columnwidth]{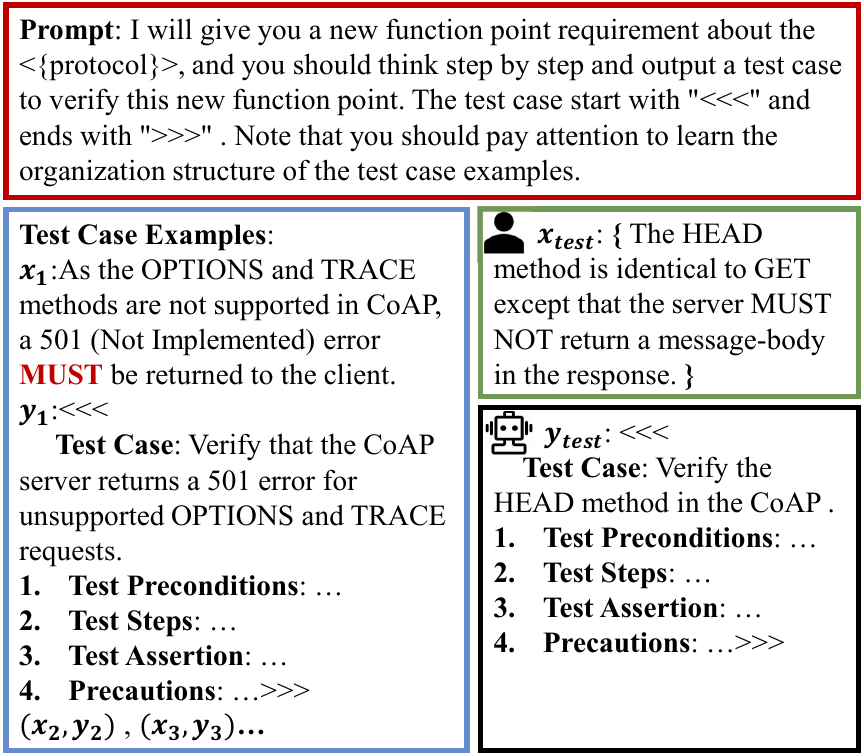}}
    \caption{Example of generating test cases using few-shot in-context learning. The contents in the red, blue, green, and black boxes represent the guidance, example, input functional point, and output test case, respectively.}
    \label{fig:TCG}
\end{figure}

Based on these functional points, keyword-based TCG calls the LLM to generate test cases.
To ensure a standardized format for the test cases, we introduce few-shot in-context learning into the LLM.
Specifically, we construct a prompt $p$ with input-output example $\{\langle x_i, y_i\rangle\}_{i=1}^{k}$, where $x_i$ represents a functional point, and $y_i$ is a standardized test case.
These input-output pairs are concatenated in the format:$\{\langle x_1, y_1 \rangle, \langle x_2, y_2 \rangle, \dots , \langle x_k, y_k \rangle\}$.
During reasoning, the test $ x_{test} $ is appended to the prompt, and the LLM learns the structure from the provided examples, generating an output $y_{test}$ in the same format, as illustrated in Fig.~\ref{fig:TCG}.
\color{black}
The test cases generated by the LLM comprise four components: test preconditions, test steps, test assertions, and precautions. The test steps describe in detail the procedure for testing the selected functional point, and the test assertions define the criteria for judging test success, which serve as the basis for evaluating the results.
\color{black}
To ensure the test cases are both domain-relevant and effective, we adopt standard processes of data generation~\cite{p_data_guo2022automated},
introducing a filter to review the functional points and their corresponding test cases, removing any anomalous test cases.

\textbf{Remark}: It is important to note that the test case generation module is optional, meaning that test cases do not necessarily have to be generated by this module. 
iPanda allows users to import their own test cases, meanwhile still providing compatibility with subsequent program generation and conformance verification functionalities. 
The main purposes of the test case generation module are twofold: to facilitate the rapid generation of effective test cases directly from protocol documents, reducing the workload for users; and to provide standardized experimental datasets for this work, enabling a quantitative evaluation of iPanda's performance.

\subsection{Automated Program Synthesis}
\label{sec:3.2}

Although LLMs have demonstrated exceptional performance, their outputs are not entirely reliable. 
For conformance testing, even when restricting the scope to a certain protocol implementation library, LLMs still exhibit the following issues during generating test programs:
\squishlist
    \item \textbf{LLM's limited understanding of code libraries}.
    Firstly, LLM's training data is inherently outdated. As Tab.~\ref{tab:cutoff} shows, even recent models have knowledge cutoffs several months in the past, leaving them unaware of new libraries. Secondly, the scarcity of open-source examples for implementation libraries also results in the lack of training data, hindering effective code generation. This phenomenon will be further demonstrated in subsequent experiments.
    \item \textbf{Hallucinations in LLM-generated code}.
    LLM hallucinations refer to instances where the LLM generates outputs that appear realistic and credible but are actually incorrect, fabricated, or baseless. This issue, an inherent artifact of the model's probabilistic nature, is difficult to completely eliminate. In this work, we define code hallucinations as generated code that: (1) calls non-existent library classes or attributes, (2) uses classes or methods incorrectly, or (3) configures required parameters improperly.
\squishend
These issues lead to inefficiencies when blindly relying on LLMs to generate task-specific programs based on a given implementation library.

\begin{table}[tbp]
  \centering
  \caption{LLM training data cutoff time\tnote{1}}
      \begin{threeparttable}
        \begin{tabular}{lll}
        \toprule
        \textbf{LLM}   & \textbf{Release date} &  \textbf{cutoff} \\
        \midrule
        GPT-o3-pro~\cite{p_more_gpt-o3-pro} & 2025/6/10 & 2024/6/1 \\
        GPT-4o(2024-11-20)~\cite{p_gpt4o_hurst2024gpt} & 2024/11/20 & 2023/10 \\
        Deepseek-R1-0528~\cite{p_more_deepseekr1} & 2025/5/28 & unknown \\
        Deepseek-V3-0324~\cite{p_more_deepseekv3} & 2025/3/25 & unknown \\
        Claude-opus-4~\cite{p_more_claude-opus-4} & 2025/5/14 & 2025/3 \\
        Gemini-2.5-pro~\cite{p_more_gemini} & 2025/6/17 & 2025/1 \\
        Qwen2.5-Coder-32B~\cite{p_qwencoder_hui2024qwen2} & 2024/11/6 & unknown \\
        \bottomrule
        \end{tabular}%
        \begin{tablenotes}
            \footnotesize \item[1] The statistics were collected on 2025/7/15.
        \end{tablenotes}
      \end{threeparttable}
  \label{tab:cutoff}%
\end{table}%
\color{black}
To address these issues, we first introduce code-oriented RAG into the long-term memory repository.
We collect all code and examples files related to the implementation.
These files naturally contain high-quality, protocol-specific knowledge that LLMs may lack.
iPanda then uses OpenAI’s text embedding model to convert these code files into semantic vector representations and builds a local vector database using Chroma~\cite{p_text-embedding-3-large}.
When generating test programs for the test cases, the long-term memory repository first retrieves code files with high similarity to the given test case based on cosine similarity. 
These retrieved results serve as contexts to help the LLM better understand how to use the implementation.

Moreover, to enable the LLM to generate high-quality test programs, we design a customized CoT method to guide the LLM in the reasoning module to perform step-by-step reasoning~\cite{p_autoiot_shen2025autoiot}. 
This process simulates how a human developer incrementally develops test subprograms, thereby enhancing the reasoning module’s capability to handle complex test cases.

\textbf{CoT1: test case understanding.} 
The LLM first decomposes the test case into multiple subtasks based on the test preconditions specified in the case. 
Specifically, for a test case that involves the collaboration of multiple participants (e.g., clients and servers), the LLM derives a clear and ordered subtask for each participant role and determines the required number of instances for each role. 
Each role instance comprises a sequence of atomic operations (e.g., connecting to the server rather than connecting and sending data).

\textbf{CoT2: subprogram generation.} Given an instance subtask, the LLM is required to generate the corresponding subprogram, strictly aligning with the subtask operations. 
The LLM uses the contexts by RAG to understand the implementation and then defines the specific behavior of each role instance in the subprogram. 
All subprograms are encapsulated as independent executable scripts and stored locally.

\textbf{CoT3: determining the subprogram execution order.} 
This step orchestrates the independent subprograms into a complete test execution flow. 
Although each role instance is independent, there are causal dependencies among them. 
For example, a server cannot accept a connection before a client requests it; a server typically starts before the client and remains in a listening state. 
Therefore, after all subprograms have been generated, the LLM infers the correct startup sequence based on the test steps defined in the test case.

\textbf{CoT4: program integration.} 
Once the startup order of the subprograms has been determined, the LLM leverages its in-context learning capability to integrate all subprograms and their corresponding execution order into a single, structured execution blueprint. 
This blueprint is output as a JSON string, which offers high readability, is easy to parse, and ensures cross-platform compatibility. 
\color{black}

\subsection{Optimizing Test Programs}
\label{sec:3.3}
\textbf{Automated Test Execution.}
When conducting automated tests, once the reasoning module generates the integrated program, testers typically need to manually launch all subprograms sequentially. 
This frequent user intervention is both cumbersome and time-consuming. 
To address this issue, we develop an automated testing platform within the execution module. 
This platform automatically parses program texts in JSON format and adaptively launches subprograms in an isolated virtual environment, ensuring that the subprograms are executed safely and under control.
If an execution error occurs (e.g. missing attributes and syntax errors), iPanda embeds the execution log into the prompt and initiates its iterative optimization mechanism to debug the program code until the generated program executes successfully. 
Once the program runs without errors, the LLM evaluates the success of the test by comparing the test assertions defined in the test case with the executed program and its results.

\textbf{Iterative Optimization Algorithm.}
Error logs encountered during program execution serve as a naturally accessible and high-quality source of knowledge.
They can effectively guide the LLM in refining and debugging.
Therefore, iPanda utilizes automated testing platform to iteratively validate and refine its output through human-like interaction~\cite{p_CRITIC_alinezhad2019critic}.
This process continues until either executable program is generated or the maximum number of retries is reached.

Specifically, given the reasoning module's LLM model $\mathcal{M}$ and the input test case $x$, the initial program solution $\hat{y}_0$ is generated by the prompt $\wp$ as $\hat{y}_0 \sim \mathbb{P}_{\mathcal{M}}(\cdot | \wp + x + r_0)$. 
The automated testing platform $\mathcal{T}$ then tests this initial solution, producing feedback $c_0 = \mathcal{T} (\hat{y}_0)$. 
For the $(i+1)$-th iteration ($i \geq 0$), the LLM's output iterates as:
\begin{equation}
    \begin{aligned}
            \hat{y}_{i+1} \sim \mathbb{P}_{\mathcal{M}} \Big( \wp & + x\\
            & + ( r_{i-m+1} + \hat{y}_{i-m+1} + c_{i-m+1} + \wp')\\
            & + ( r_{i-m+2} + \hat{y}_{i-m+2} + c_{i-m+2} + \wp')\\
            & +  \dots\\
            & + ( r_{i} + \hat{y}_{i} + c_{i} + \wp') \\
            & + r_{i+1} \Big),
    \end{aligned}
\end{equation}
where $\wp'$ is a prompt template designed to guide the LLM in debugging. 
The parameter $1 \leq m \leq i+1$ represents the window size of the short-term memory cache, preventing the input length from exceeding the maximum text length that the LLM can handle (e.g., GPT-4o supports up to 128K tokens~\cite{p_gpt4o_hurst2024gpt}). 
When $m=i+1$, it considers all historical interaction information.
The $r_i$ is the context retrieved from long-term memory repository by $\mathcal{R}$, specifically as:
\begin{equation}
    r_i = 
    \begin{cases}
        \mathcal{R} ( \wp + x ), & i = 0,\\
        \mathcal{R} ( c_{i-1} + \wp'), & i \geq 1. 
    \end{cases}
\end{equation}
This optimization process continues until the generated program meets specific execution success criteria or the maximum step of iterations is reached. 
The pseudocode for iterative optimization algorithm is shown in Alg.~\ref{alg:cirtic}.
\begin{algorithm}[!ht]
	\caption{Iterative Optimization Algorithm}
    \label{alg:cirtic}
	\begin{algorithmic}[1]
        \REQUIRE
          Input $x$, initial prompt $\wp$, iteration prompt $\wp'$, LLM $\mathcal{M}$, testing platform $\mathcal{T}$, retriever $\mathcal{R}$, maximum step $n$
        \ENSURE
          output $\hat{y}$ from $\mathcal{M}$
        \STATE $r \leftarrow \mathcal{R}(\wp + x)$
        \STATE $seq \leftarrow \wp + x + r$
        \STATE Generate initial output $\hat{y}_0 \sim \mathbb{P}_\mathcal{M}(\cdot | seq)$
        \FOR{$i \leftarrow 0$  to $i\leftarrow n-1$}
            \STATE $c \leftarrow \mathcal{T}(\hat{y}_i)$
            \IF{$c$ indicates that $\hat{y}_i$ is correct }
                \STATE \textbf{return} $\hat{y}_i$
            \ENDIF
            \STATE $r \leftarrow \mathcal{R}(\wp' + c)$
            \STATE $seq \leftarrow seq + \hat{y}_i + c + \wp' + r$
            \STATE Generate $(i+1)$-th output $\hat{y}_{i+1} \sim \mathbb{P}_\mathcal{M}(\cdot | seq)$
        \ENDFOR
        \STATE \textbf{return} $\hat{y}_n$
	\end{algorithmic}
\end{algorithm}

\textbf{Remark:} During the iteration, the test program is continuously refined.
As the number of iterations increases, the length of the input context grows linearly, which in turn increases the reasoning load on the LLM. 
Therefore, it is crucial to set an appropriate cache window size to limit the context length.
Additionally, the more iterations, the smaller the marginal benefit of program correction.
Hence, selecting an optimal maximum step of iterations is essential. 
Based on experimental observations, we set the default maximum step to $6$.

\subsection{Summarizing Bugs}
\label{sec:3.4}
The summarization module performs two key functions:
\squishlist
    \item \textbf{Summarizing program generation experience}.
    With appropriate prompt guidance, the module generates experience $\hat{s}$ by analyzing the program testing process 
    $\wp + x + (\hat{y}_0 + c_0) + \dots + (\hat{y}_{t-1} + c_{t-1}) + \hat{y}_t$.
    The experience is stored in the long-term memory repository as valuable knowledge to continuously enhance iPanda’s overall performance.
    By leveraging the accumulated experience, the iterative optimization algorithm can refine program corrections more efficiently, ensuring that iPanda continuously learns and improves its ability to handle similar tasks in the future with greater efficiency and accuracy.
    \item \textbf{Ensuring conformance between protocol and its implementation}.
    With prompts guidance, the module also reviews the test cases and the usage of the protocol implementation.
    It evaluates whether the tested protocol implementation adheres to the specifications embedded in the test cases, and generates corresponding conformance testing reports.
\squishend

A primary challenge is distinguishing genuine conformance testing results from incidental code generation errors. Due to inherent randomness and the imperfect adherence to instructions, the LLM may erroneously report on its own code generation errors (e.g., incorrect parameter passing or invalid attribute calls) instead of actual conformance violations. To isolate the true conformance test results, we implement a two-stage filtering process. Firstly, a keyword-based filter automatically removes reports containing terms associated with common generation bugs,  including incorrect parameter passing (e.g., incorrectly binding to any-address) and incorrect method or attribute calling (e.g., some attribute does not exist). Subsequently, the remaining reports undergo manual review to verify their relevance and accuracy.

%% file: 4_experiment.tex
\section{Evaluation}
\label{sec:4}
In the section, we present our experimental setup, results and analyses, to demonstrate iPanda's effectiveness.

\subsection{Experimental Setup}
\label{sec:4.1}

\subsubsection{Experimental platform}
\label{sec:4.1.1}
We implemented the iPanda using Python and deployed it on our local server.
All target protocol implementations, along with their required libraries, were pre-installed in a local virtual environment.
For LLM, we used GPT-4o~\cite{p_gpt4o_hurst2024gpt}, DeepSeek-V3~\cite{p_deepseekv3_liu2024deepseek}, and Qwen2.5-Coder-32B~\cite{p_qwencoder_hui2024qwen2}. 
These models represent the most advanced base models, differing in architecture, size, and pretraining focus. 

All LLMs were configured with short-term memory window size $m=10$, $temperature$ $=$ $0$, and $top\_p$ $=$ $0.1$. 
For RAG, we utilized OpenAI’s text-embedding-3-large as embedding model~\cite{p_text-embedding-3-large}. 
For simplicity and generality, the testing environment was configured locally, using the local IP address and different ports to simulate network conditions.

\subsubsection{Tested protocols and Implementations}
\label{sec:4.1.2}
We selected the following protocols and their Python implementations:
\squishlist
    \item \textbf{CoAP \& \texttt{aiocoap}}~\cite{p_aiocoap}. The Constrained Application Protocol (CoAP) is a lightweight network communication protocol designed for resource-constrained devices, primarily used for IoT devices. 
    It was standardized in 2014 as RFC 7252~\cite{p_coap_rfc7252_shelby2014rfc}. 
    As of March 2025, there are $32$ RFC documents related to CoAP~\cite{p_coap_32}.
    We selected \texttt{aiocoap}, the Python implementation of CoAP, as the target for conformance testing. 
    As of March 2025, \texttt{aiocoap} fully or partially supports $9$ CoAP-related RFC documents.
    \item \textbf{RSocket \& \texttt{rsocket-py}}.
    RSocket is an application protocol that provides reactive stream semantics over an asynchronous, binary boundary. 
    As of March 2025, RSocket has not yet been formalized as a RFC standard. 
    However, it has a well-established specification~\cite{p_rsocket} (adhering to RFC 2119) and implementations across various programming languages. 
    We selected \texttt{rsocket-py}, the Python implementation of RSocket, as the target for conformance testing.
\squishend
The key reason for selecting them is that they meet important criteria for conformance testing: (1) the protocols have established mature specifications, but their implementations have only recently developed; (2) only partial specifications have been implemented in the implementations; (3) there are no mature conformance testing tools for these implementations.


Notably, LLMs possess varying levels of pre-existing knowledge about specific protocols. Therefore, the RAG was adapted based on the LLM's prior knowledge of each library. A preliminary assessment revealed that the LLM has extensive built-in knowledge of \texttt{aiocoap}, so we deactivated the long-term memory repository to isolate its inherent capabilities. Conversely, for \texttt{rsocket-py}, a more recent library with limited open-source examples, the LLM lacked sufficient knowledge. Therefore, we retained the long-term memory repository and augmented it with the library's source code to provide the necessary context.


\subsubsection{Dataset}
\label{sec:4.1.3}
Using the keyword-based TCG in Sec.~\ref{sec:3.1}, we introduced two test case sets for CoAP and RSocket:
\squishlist
    \item \textbf{CoAP-set}. 
    We selected $11$ RFC documents directly related to CoAP to generate the CoAP-set, which consists of $231$ uniformly formatted test cases.
    \item \textbf{RSocket-set}. 
    We generated the RSocket-set using RSocket document, having $62$ test cases.
\squishend
All test cases underwent professional manual review and selection to ensure their authenticity and validity.
Each test case follows a standardized format comprising five components: test case name, test preconditions, test steps, test assertions, and precautions.
Notably, the core of our work is on the accuracy of iPanda’s code generation and the effectiveness of its conformance testing. 
Therefore, rather than pursuing exhaustive specification coverage, our work prioritizes the validity and relevance of individual test cases. To ensure the consistency and comparability of our findings, we employ the same test cases across all experiments.

\subsubsection{Baseline}
\label{sec:4.1.4}
Specification-driven conformance testing methods typically require the development of protocol-specific conformance testing tools. 
Unfortunately, no open-source conformance testing tools currently exist for CoAP and RSocket. 
Moreover, our work is the first to employ LLMs for conformance testing in a protocol-agnostic manner.
To reasonably evaluate iPanda’s performance, we selected a pure-LLM baseline approach in which the LLM generates test program in a single step (using the same prompt and setup). 
This corresponds to the startup stage of iPanda with the memory module disabled, serving as our experimental baseline.

\subsubsection{Metrics}
\label{sec:4.1.5}
$\textbf{Pass@k}$.
This metric is commonly used to evaluate the correctness and reliability in generating code for a given programming task, which is a core sub-task of iPanda.
$Pass@k$ measures the probability that at least one of the $k$ generated code solutions is correct.
Specifically, for a test set containing $N$ cases $\{T_1, T_2, \dots, T_N\}$, the $Pass@k$ metric for an LLM $\mathcal{M}$ is defined as:
\begin{equation}
    Pass@k = \frac{\sum\limits_{i=1}^{N} \left( \phi_1(T_i) \oplus \phi_2(T_i) \oplus \dots \oplus \phi_k(T_i) \right)}{N},
\end{equation}
where $\oplus$ denotes the XOR operator, and $\phi_j(\cdot) \in \{0,1\}$ represents the evaluation of the $j$-th generated solution.
It takes a value of $1$ if the solution is correct and $0$ otherwise.

\textbf{Conformance testing results}. 
To evaluate the effectiveness of iPanda in conformance testing, we conduct a qualitative assessment of the generated programs compliance with specifications. The assessment categorizes results into two types:
\squishlist
    \item \textbf{Positive samples}: The generated program is executable and functions in accordance with the specification.
    \item \textbf{Negative samples}: The code fails to execute or functionally violates the protocol requirements.
\squishend

\subsection{Marginal Benefit Analysis}
\label{sec:4.2}

\begin{figure*}[htbp]
    \centering{\includegraphics[width=1.95\columnwidth]{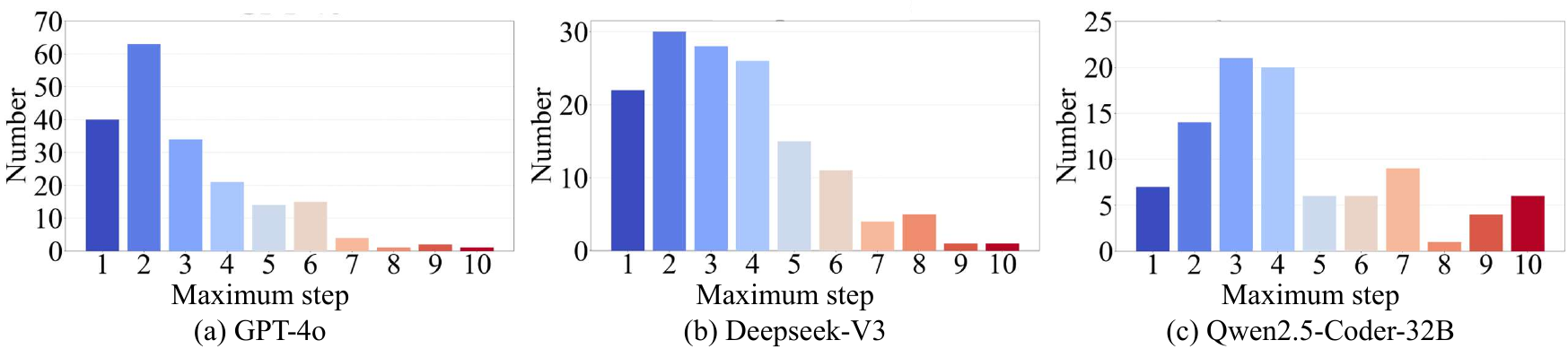}}
    \caption{The number of code successfully generated within the maximum iteration step limit, using (a) GPT-4o, (b) Deepseek-V3, and (c) Qwen2.5B-Coder-32B.}
    \label{fig:performance}
\end{figure*}

As the number of iterations increases, the marginal benefit of program correction gradually decreases. 
To achieve the optimal marginal benefit, we must first determine an appropriate maximum step of LLM iterative reasoning. 
In our experiments, we set the upper limit for reasoning iterations to $10$. 
iPanda was tested on the CoAP-set, which contains more test samples. 
To analyze the required number of reasoning iterations for successfully generating executable program, we constructed a histogram depicting the cost of successful generations. 
The experimental results are shown in Fig.~\ref{fig:performance}.
It can be observed that in most cases, no more than $6$ steps are required to generate executable program. 
Specifically, when setting the maximum step to $10$, using GPT-4o, iPanda successfully tested $195$ cases in the CoAP-set (classified as positive samples), among which $187$ cases required no more than $6$ iterations, accounting for $187/195 = 95.90\%$. 
Using DeepSeek-V3, the corresponding proportion was $132/143 = 92.31\%$.
Using Qwen2.5-Coder-32B, the proportion was $74/94 = 78.72\%$.
Under the influence of the scaling law, large-scale foundation models such as GPT-4o and DeepSeek-V3 (both exceeding $100 B$ parameters) exhibit stable performance in program generation and correction. 
As the number of iterations increases, both marginal benefit steadily declines.
However, for Qwen2.5-Coder-32B, which has only $32 B$ parameters, the marginal benefit is relatively unstable. 
This may be attributed to the model’s smaller scale, leading to weaker program generation capabilities.
Resultly, we set the maximum step of iterations to $6$ in subsequent experiments to achieve the optimal marginal benefit.

\subsection{Performance Analysis of Program Generation}
\label{sec:4.3}
We conducted experiments using the baseline and iPanda supported by the three LLMs to evaluate their code generation performance on the CoAP-set. 
The $Pass@1$ experimental results for all methods are shown in Tab.~\ref{tab:comparison}.
Experimental results show that under the same LLM conditions, iPanda significantly enhances program generation capabilities compared to the baseline. 
$Pass@1$ is improved by $4.675\times$(GPT-4o), $6\times$(Deepseek-V3), and $10.751\times$(Qwen2.5-Coder-32B) respectively.
For example, when using the GPT-4o, the baseline achieves only a $Pass@1$ of $17.32\%$.
In contrast, iPanda improves this by $4.675\times$, reaching $80.95\%$.
This strongly validates that \texttt{aiocoap} aligns with the specifications underlying these test cases, reducing the likelihood of nonconformances and narrowing the focus for critical testing.
Even with the smallest model Qwen2.5-Coder-32B, iPanda's $Pass@1$ is nearly twice that of the baseline using the most advanced model GPT-4o. 
These results confirm the success of the iterative optimization algorithm. 
With its support, even smaller-scale LLMs can be stimulated to exhibit stronger program generation capabilities, achieving performance comparable to or even surpassing larger, more advanced LLMs.

\begin{table}[htbp]
  \centering
  \caption{$Pass@1$ comparison on CoAP-set}
    \begin{tabular}{lll}
    \toprule
    \textbf{Used-LLM} & \textbf{Baseline} & \textbf{iPanda} \\
    \midrule
    GPT-4o & 17.32\% & \textbf{80.95\%} \\
    Deepseek-V3 & 9.52\% & \textbf{57.14\%} \\
    Qwen2.5-Coder-32B & 3.03\% & \textbf{32.03\%} \\
    \bottomrule
    \end{tabular}%
  \label{tab:comparison}%
\end{table}%

\begin{figure}[htbp]
    \centering{\includegraphics[width=1\columnwidth]{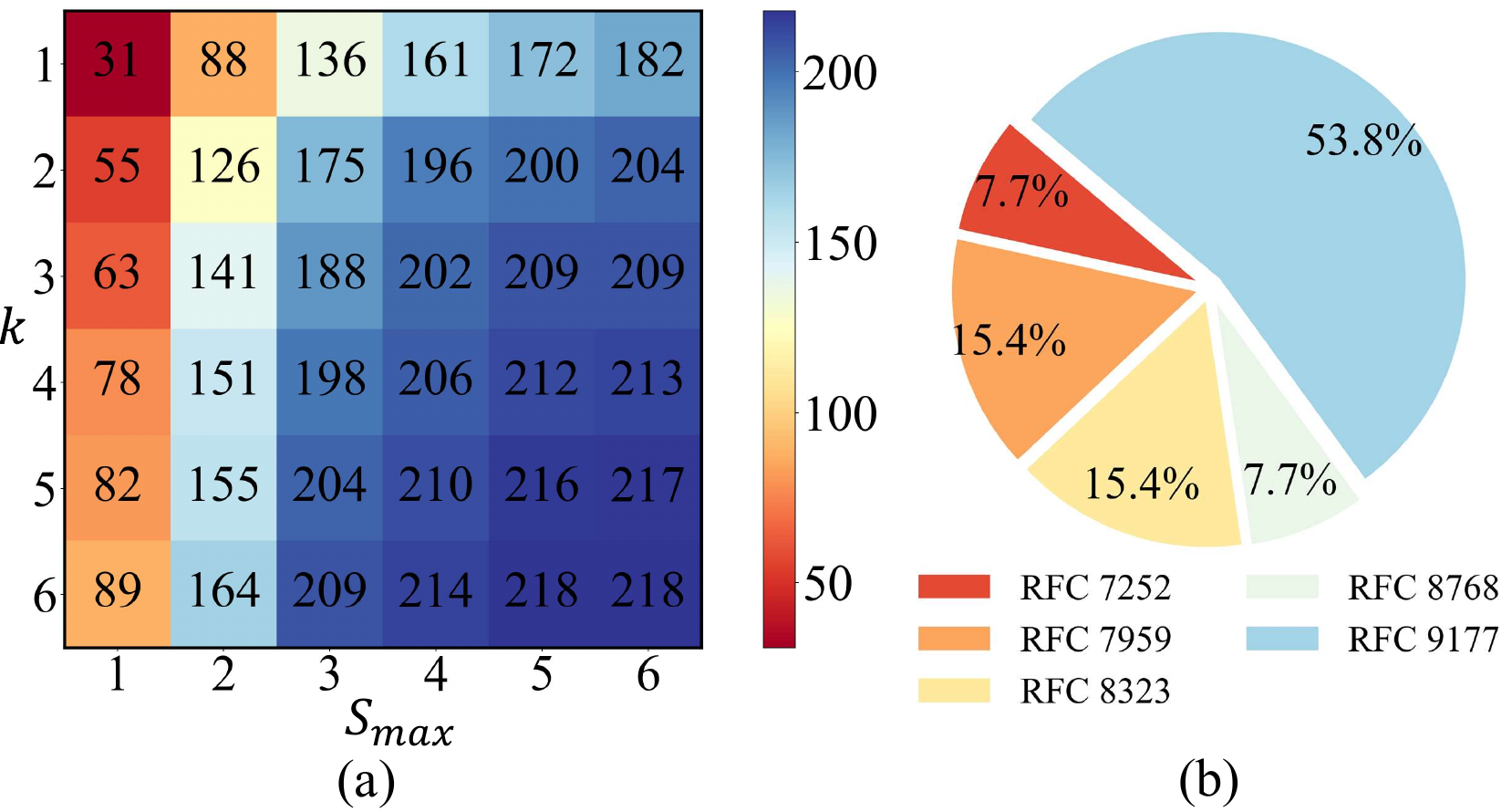}}
    \caption{(a) The joint experiment of maximum reasoning step $S_{max}$ and repetition numbers $k$. The numbers in squares represent the number of executable program generated by iPanda under the $S_{max}$ and $k$.   (b) Statistics of RFC documents containing failed test cases under $S_{max}=6$ and $k=6$. The size of the sector reflects how many failed test cases the RFC document contains. }
    \label{fig:step_k_and_negetive_sample}
\end{figure}

Due to the inherent randomness in LLM-generated content, multiple repeated tests are typically required for program generation, and performance is often evaluated using $Pass@k$. 
Therefore, we conducted a joint experiment to examine the impact of the maximum reasoning iteration step $S_{\max}$ and the number of repeated tests $k$.
Given the strong performance of GPT-4o, we selected it as the default LLM for iPanda. 
The experiment was configured with $S_{\max} = \{1,2,3,4,5,6\}$ and $k = \{1,2,3,4,5,6\}$. 
The experimental results on the CoAP-set are shown in Fig.~\ref{fig:step_k_and_negetive_sample}-(a), where the numbers in squares represent the number of executable program generated by iPanda under the current $S_{max}$ and $k$. 
When setting $S_{\max} = 6$ and $k = 1$, the experiment degenerates into the setup of Fig.~\ref{fig:performance}-(a).
When setting $S_{\max} = 1$ and $k = 6$, the experiment degenerates into the baseline $Pass@6$ evaluation.
From the results, we observe that increasing the reasoning iteration step significantly improves iPanda’s performance. 
Additionally, repeated testing can also enhance the success rate of program generation, though its impact is considerably smaller than increasing reasoning iterations. 
For instance, increasing the number of repeated tests by five (i.e., $k=6$ instead of $k=1$) raises the number of positive samples from $31$ to $89$.
Increasing the number of inference iterations by five (i.e., $S_{\max}=6$ instead of $S_{\max}=1$) raises the number of positive samples from $31$ to $182$.
These results validate the rationality of our decision to adopt a sequential strategy (i.e., iterative program generation using a single LLM), rather than a parallel strategy (i.e., involving multiple LLMs generating program simultaneously).
While the former has a higher interaction cost per step due to increased context, this additional context contains valuable execution feedback, providing richer information to guide the LLM in refining the generated program.

\subsection{Ablation Study}
\label{sec:4.4}
To validate the effectiveness of the methods used in iPanda, we conducted an ablation study to analyze the contributions of code-oriented RAG and iterative optimization algorithm. 
The experimental results are presented in Tab.~\ref{tab:ablation}.
\begin{table}[tbp]
  \centering
  \caption{Ablation study on $Pass@1$}
    \begin{tabular}{lll}
    \toprule
    \textbf{Approach} & \textbf{CoAP-set} & \textbf{RSocket-set} \\
    \midrule
    iPanda & \textbf{80.95\%} & \textbf{38.71\%} \\
    iPanda, w/o code-oriented RAG & 80.95\% & 14.51\% \\
    iPanda, w/o iterative optimization & 17.32\% & 3.23\% \\
    Baseline & 17.32\% & 11.29\% \\
    \bottomrule
    \end{tabular}%
  \label{tab:ablation}%
\end{table}%
Since LLMs have a deep understanding of CoAP and \texttt{aiocoap}, the long-term memory repository was frozen, disabling the code-oriented RAG. As a result, RAG does not impact the testing outcomes for this protocol, leading to only two experimental configurations in the ablation study.
To evaluate the effectiveness of code-oriented RAG, we introduced additional tests for the RSocket protocol and its Python implementation, \texttt{rsocket-py}. In this setup, iPanda was preloaded with the open-source code of \texttt{rsocket-py} for use in RAG.
Experimental results on the RSocket-set revealed that the inclusion of code-oriented RAG significantly improved LLM’s ability to understand and correctly utilize \texttt{rsocket-py}. 
This resulted in an increase in $Pass@1$ from $14.51\%$ to $38.71\%$.

When the iterative optimization algorithm is removed, iPanda degrades into the baseline that only incorporates code-oriented RAG. 
Experimental results show that, on both datasets, the performance of iPanda drops significantly without the iterative optimization algorithm. 
This phenomenon has been extensively studied in previous experiments.
Nevertheless, on the RSocket-set, even without the iterative optimization algorithm, iPanda still outperforms the baseline. 
This further demonstrates the effectiveness of code-oriented RAG.

\textbf{Remark:} The primary goal of iPanda is to support conformance testing for multiple communication protocols. 
Therefore, it must exhibit protocol compatibility.
Meanwhile, we also expect iPanda’s performance to improve as the underlying LLMs advance, requiring it to maintain model compatibility.
The extensive performance and ablation experiments conducted above confirm that iPanda demonstrates strong compatibility with both different protocols and different LLMs.

\subsection{Results Analysis of Conformance Testing}
\label{sec:4.5}
According to the joint experiment results in Sec.~\ref{sec:4.3}, maximizing the number of iterations and repetitions yields the highest number of positive samples, indicating that \texttt{aiocoap} conforms to the majority of protocol specifications. 
For negative samples, we analyzed the documents they belong to, as shown in Fig.~\ref{fig:step_k_and_negetive_sample}-(b). 
Notably, Most of the negative samples are concentrated in the CoAP's RFC 9177. 
This suggests that \texttt{aiocoap} has likely not yet implemented the specifications outlined in RFC 9177.
This hypothesis aligns with the author's introduction on GitHub~\cite{p_aiocoap}, which specifies the standards supported by \texttt{aiocoap}, confirming that RFC 9177 has not yet been implemented in \texttt{aiocoap}.
By generating as many positive samples as possible, iPanda can effectively reduce the scope of conformance testing. In addition, the test results of negative samples also guide further testing.

%% file: 5_relatedWork.tex
\section{Related work}
\label{sec:5}

\textbf{Traditional testing tools}.  
Several tools have been developed for protocol testing. Testing and Test Control Notation version 3 (TTCN-3) is a widely used language for rigorous conformance certification in mobile communications, and IoT protocols~\cite{p_ttcn-3_grabowski2003introduction}. Scapy enables flexible construction and transmission of custom packets via scripting, aiding tests of protocol edge cases and exception handling, particularly in security assessments~\cite{p_scopy_rohith2018scapy}. Protocol fuzz testing tools, such as Fairfuzz and Boofuzz, have demonstrated strong capabilities in uncovering vulnerabilities and testing robustness~\cite{p_boofuzz_pereyda2019boofuzz, p_more_fairfuzz_lemieux2018fairfuzz, p_more_t-fuzz_peng2018t}. However, these tools are typically applied at specific stages within existing testing workflows and heuristic-based approaches~\cite{p_testshark_makhmudov2025online}. A major limitation remains their dependence on the manual creation of test cases and scripts, requiring substantial developer effort. Currently, no tool seamlessly integrates protocol documentation, implementation, and tests into one conformance testing process. Addressing this gap is precisely the focus of our work.

\textbf{LLM-based automated testing tools}.  
Automated testing tools based on LLMs have gained increasing attention in recent years. Researchers have begun exploring LLM-powered agents for automating testing tasks. In code testing, some studies leveraged LLMs to generate high-coverage unit tests, such as those for the JUnit testing framework~\cite{p_jnuit_siddiq2024using, p_more_junit_guilherme2023initial}. 
PENTESTGPT, an penetration testing tool, effectively identifies common vulnerabilities and analyzes source code for flaws~\cite{p_agent-survey-128_pentestgpt_deng2023pentestgpt}. DB-GPT integrates the Tree-of-Thought method into LLMs to systematically analyze database anomalies~\cite{p_agent-survey-41_dbgpt_zhou2023llm}. In fuzz testing, existing tools often struggle with message format obfuscation and dependencies between messages. To address this, LLMIF~\cite{p_llmif_wang2024llmif} integrates LLMs into IoT fuzz testing to automate the extraction of protocol formats and device response inference. However, those LLM-driven testing tools typically target isolated tasks and do not effectively support conformance testing. 
\color{black}
RFCScan is most similar to our work.
It builds semantic indexes for both implementation code and specifications to detect conformances between the code and the RFC specifications at the semantic level. However, this approach focuses on static comparisons between code and specifications, rather than simulating how developers write test cases to verify the actual runtime behavior of the protocol, which is precisely where our work differs.
\color{black}
To address this research gap, we introduce iPanda, an LLM-based agent tailored specifically for protocol conformance testing.

\textbf{Augmented LLMs}.  
Although LLMs excel in tasks such as question answering and text generation, their performance remains constrained by the limitations of pre-trained datasets and the context provided during inference. 
As a result, they may perform poorly in certain specialized tasks. 
To address these limitations, researchers have explored various tools to enhance LLM capabilities, including Web browsers~\cite{p_autodroid-8_deng2023mind2web, p_autodroid-27_nakano2021webgpt, p_more_web_chen2025weinfer}, RAG~\cite{p_more_rag_du2024vul, p_more_rag_jeong2023generative, p_more_rag_ng2025rag}, programming tools~\cite{p_CRITIC_alinezhad2019critic, p_more_code_lu2023llama}, other deep neural network models~\cite{p_autodroid-32_shen2023hugginggpt}, and so on~\cite{p_more_selfconsistency_wang2022self, p_more_react_yao2023react, p_more_reflexion_shinn2023reflexion}.  
Similarly, in our work, iPanda incorporates RAG and programming tools to enhance the effectiveness of LLMs, improving their ability to analyze protocol specifications, generate test cases, and interact with protocol implementation libraries.

%% file: 6_conclusion.tex
\section{Conclusion}
\label{sec:6}

We present iPanda, the first LLM-based intelligent agent for automated  conformance testing of communication protocols.
iPanda leverages the keyword-based TCG method to efficiently generate test cases from protocol documents, and employs code-oriented RAG and customized CoT strategy to enhance its understanding of tested protocol implementation and generate test program.
Furthermore, the ierative optimization algorithm is integrated for iterative program refinement in automated testing platform.  
Through execution-based validation, iPanda assesses whether the protocol implementation adheres to the specified protocol requirements.
Experimental results prove the effectiveness and high efficiency of iPanda in conformance testing.